\documentclass[global,twocolumn]{svjour}
\usepackage{graphics}
\usepackage{sistyle}
\usepackage{longtable,tabularx,hhline}

\journalname{Applied Physics B}
\newcommand{\ion}[2]{\mbox{$^{#2}$#1$^+$}}
\newcommand{\Ca}[1]{\ion{Ca}{#1}}

\newcommand{\lev}[3]{\mbox{$^{#1}$#2$_{\mbox{\tiny$#3$}}$}}

\newcommand{\Isat}{\mbox{$I_{\mathrm{sat}}$}}

\newcommand{\unit}[1]{\,\mbox{#1}}

\newcommand{\uW}{\unit{$\mu$W}}

\newcommand{\nm}{\unit{nm}}

\newcommand{\persec}{\unit{s$^{-1}$}}
\newcommand{\ms}{\unit{ms}}

\newcommand{\ish}{\mbox{$\sim$}\,}

\begin{document}
%
\title{Background-free detection of trapped ions}
\author{N. M. Linke \and D. T. C. Allcock \and D. J. Szwer\thanks{\emph{Present address:} Department of Physics, University of Durham, U.K.} \and C. J. Ballance \and T. P. Harty \and H. A. Janacek \and D. N. Stacey \and A. M. Steane \and D. M. Lucas
%
}                     

%
%
\institute{Department of Physics, University of Oxford, Parks Road, Oxford OX1 3PU, U.K.}
\date{Received: date / Revised version: date}
%
\maketitle
\begin{abstract}
We demonstrate a Doppler cooling and detection scheme for ions with low-lying D levels which almost entirely suppresses scattered laser light background, while retaining a high fluorescence signal and 
efficient cooling. We cool a single ion with a laser on the $\lev{2}{S}{1/2}\leftrightarrow\lev{2}{P}{1/2}$ transition as usual, but repump via the \lev{2}{P}{3/2} level. By filtering out 
light on the cooling transition and detecting only the fluorescence from the $\lev{2}{P}{3/2}\rightarrow\lev{2}{S}{1/2}$ decays, we suppress the scattered laser light background count rate 
to 1\persec\ while maintaining a signal of 29000\persec\ with moderate saturation of the cooling transition. This scheme will be particularly useful for experiments where ions are trapped in 
close proximity to surfaces, such as the trap electrodes in microfabricated ion traps, which leads to high background scatter from the cooling beam.
\end{abstract}
%
\section{Introduction}
\label{intro}
In recent years trapped ions have been at the heart of an increasing range of experiments and proposals \cite{Wineland11,LEI03}, most prominently the field of quantum information processing (QIP). 
There has been a significant trend towards ever-smaller two- and three-dimensional trap geometries, as well as towards integration of imaging optics \cite{Brady11} and optical cavities 
\cite{Dantan10,Leibrandt09,Russo09,Keller07} with the trap electrode structure. There have also been studies of trapped ions' interactions with a thin conducting wire \cite{Daniilidis09} and proposals to extend 
this to nano-fibres \cite{BrownnuttSelber} in order to couple ions to the fibre's evanescent light field (as previously implemented with neutral atoms \cite{Sagu07,Vetsch10}). Other 
experiments pursue the entanglement of the internal state of a trapped ion with the vibrational mode of a cantilever in order to study quantum effects in meso- and macroscopic objects 
\cite{Hensinger05}. 

Many of these experiments rely on ions being trapped close to surfaces and all of them depend on laser cooling \cite{Chu85} as the standard technique to reduce the ions' motion. For its 
large velocity capture range, high cooling rates and straightforward implementation Doppler cooling is usually the initial stage of the cooling process \cite{ESH03}. In a two-level approximation the Doppler cooling 
limit temperature $T = \hbar \Gamma / 2 k_{\mathrm B}$ (where $\Gamma$ is the natural linewidth of the cooling transition) is achieved in the limit of low laser intensity. However, the 
cooling rate is greater at higher intensities, where the photon scattering rate is larger, and in practice intensities around the saturation intensity $\Isat = 2\pi h c \Gamma /\lambda^3$ 
(where $\lambda$ is the transition wavelength) offer a good compromise between these competing demands. Maintaining a high cooling efficiency is particularly important in surface-electrode 
ion traps, where the shallower trap depth can lead to ion loss after background gas collisions if the cooling rate is low and where anomalous heating effects are more prevalent than in three dimensional traps due to the proximity of the ion to the trap electrodes.

When detecting a fluorescence signal $S$ from ions trapped in close proximity on the order of \SI{100}{\mu m} or less to surfaces (the trap electrodes, integrated optics or other objects of 
study), light scattered off the surroundings is picked up by the imaging or detection optics and can significantly increase the background signal $B$. The signal is usually shot-noise 
limited, so the signal-to-noise ratio is given by $S / \sqrt{S+B}$ and reducing the background is clearly advantageous (see \cite{Burrell10} for a more detailed discussion of discrimination 
of single-ion fluorescence from background). Previous work achieved negligible background by driving the narrow dipole-forbidden $\lev{}{S}{1/2}\leftrightarrow\lev{}{D}{5/2}$ transition in 
\Ca{40} with $\sim\SI{250}{mW}$ of laser intensity and collecting fluorescence on allowed decay transitions \cite{Hendricks08}; however this had the disadvantages of reducing both the cooling and 
fluorescence rates significantly, and requiring a more complex laser system.

Here, we describe a new scheme that achieves detection at essentially zero background but only uses dipole-allowed transitions and hence does not require a high power or narrow linewidth laser. After a brief outline of the experimental setup, we describe four different repumping schemes for \Ca{40} (which are readily applicable to ions with similar level structure), implement them and compare the results. A similar method has previously been used in magneto-optical trapping of neutral $^{85}$Rb \cite{Freegarde09}.

We note that the new repumping schemes introduced here, whilst useful
for detecting the ion itself, are not applicable to detection of the
internal state of the ion by the electron-shelving technique \cite{myerson08} because the \lev{}{D}{5/2} \textquotedblleft shelf" is part of the cooling cycle. However, they could be useful in the context of ion trap QIP for example for monitoring fluorescence and maximizing the cooling rate for sympathetic cooling ions \cite{Barrett2003}. Since they concern photon detection they also have no influence on the charging effects caused by scattering light off surfaces close to the ion as described in \cite{Harlander10}.

\section{Experimental setup}
\label{sec:expsetup}

\begin{figure}
\begin{center}
\resizebox{0.45\textwidth}{!}{\includegraphics{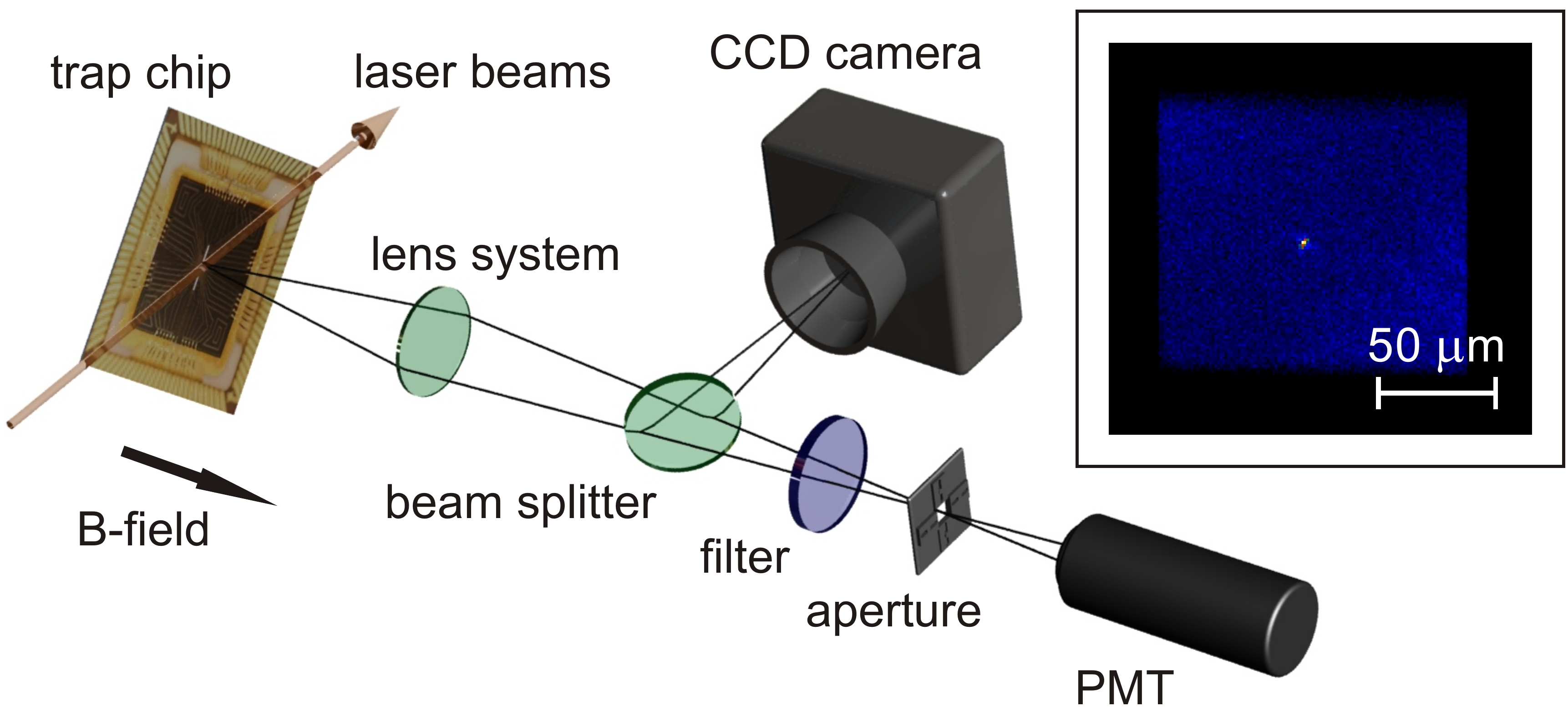}}
\end{center}
\caption{Experimental setup. The ion is trapped in a surface trap, left. All laser beams are superimposed and propagate at $45^\circ$ to the trap axis; they are linearly 
polarized parallel to the chip surface and perpendicular to the magnetic field ($\sigma^\pm$ polarization). The ion is imaged through a lens system onto an adjustable aperture and a PMT, or 
optionally via a beam-splitter onto a CCD camera. (For the image shown, the aperture was re-imaged onto the camera to show both ion and aperture.) A significant fraction of the ion 
fluorescence falls outside the sharp central image; with a diffraction-limited optical system the aperture could be stopped down to reduce still further the background scattered laser light. 
The 393\nm\ filters (see text) can be placed in front of the PMT.}
\label{fig:expsetup}       
\end{figure}

Figure \ref{fig:expsetup} shows a sketch of the trap and detection system.  Single \Ca{40} ions are trapped in a linear surface-electrode radiofrequency (RF) microchip trap fabricated at 
Sandia National Laboratories (for details please refer to \cite{Stick10} and \cite{Allcock11}). The electrode surfaces are aluminium, supported above the silicon substrate by silicon dioxide 
insulating pillars, and the electrodes overhang the pillars to shield trapped ions from the insulating surfaces. The ion is trapped \SI{84}{\mu m} above the plane of the electrodes. The 
radial confinement is provided by a harmonic RF pseudo-potential driven at \SI{33}{MHz} with an amplitude of \SI{200}{V_{pp}}, which gives a radial trap frequency of \SI{3.2}{MHz}. Further 
electrodes carry DC potentials to provide for axial confinement (axial trap frequency \SI{1.0}{MHz}), tilting of the radial principal axes, and micromotion compensation. The trap is enclosed 
in an ultra-high vacuum chamber with a residual pressure of $<1\times 10^{-11}$\,Torr. An ion is loaded isotope-selectively by photo-ionization \cite{Lucas04} from a beam of neutral calcium 
atoms provided by a thermal oven. The oven is mounted behind the trap chip, sending the atoms through a slit (width \SI{100}{\mu m}) in the trap centre. The applied magnetic field is 
approximately \SI{2.4}{G} and aligned perpendicular to the chip. 

Extended-cavity diode lasers at 397\nm, 850\nm, 854\nm\ and 866\nm\ are available to Doppler-cool and repump the ion (see level diagram in figure~\ref{fig:levelscheme}); all are locked 
to low-finesse optical cavities (drift rates $<\SI{0.5}{MHz/hour}$, laser linewidths $\approx\SI{0.5}{MHz}$). All the laser beams are superimposed and polarized linearly perpendicular to the applied magnetic field. They are 
sent across the chip parallel to the surface at an angle of $45^\circ$ to the trap axis. The principal axes of the pseudo potential are tilted using a static quadrupole field (for details 
see \cite{Allcock10}) to allow cooling of all motional modes. Spot sizes ($1/$e$^2$ intensity radius) at the ion of the 397\nm\ and infra-red beams are (21$\mu$m$\times$31$\mu$m) and (17$\mu$m$\times$22$\mu$m) 
respectively.

The ion is observed with a photomultipler\footnote{Electron Tubes P25PC} (PMT) or an electron-multiplying CCD camera\footnote{Andor Luca 285\_Mono,USB} through an imaging system with a numerical aperture of 0.25 and a 
magnification of 8. At a focal point in the imaging path, an adjustable aperture is used to reduce background scatter. For the new repumping schemes described below, scattered laser light at 
397\nm\ is filtered out using a pair of 393\nm\ interference filters\footnote{Semrock FF01-387/11-25} (with net transmission measured to be 78\% and 0.3\% at 393.5\nm\ and 397.0\nm\ 
respectively). 

\section{Repumping schemes}
\label{sec:scheme}

\begin{figure}[t]
\begin{center}
\resizebox{0.35\textwidth}{!}{\includegraphics{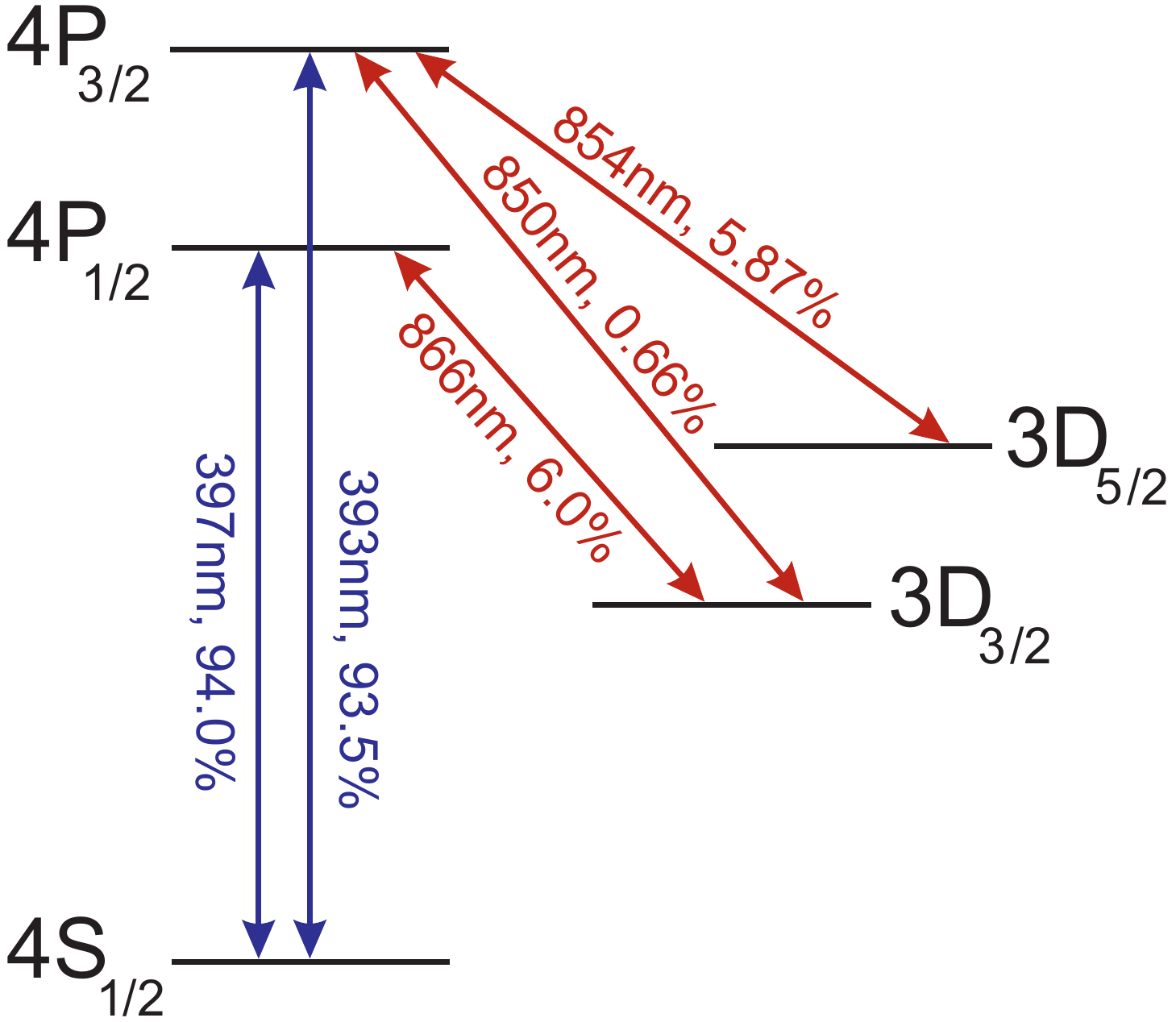}}
\end{center}
\caption{Level scheme of \Ca{40} with dipole transitions and their respective wavelengths and branching ratios \cite{Gerritsma08}. Our repumping schemes are labelled by the number of infra-red repumpers used, prefixed by ``F'' if the filter is in place and only \SI{393}{nm} light is detected. A laser on the main cooling transition at \SI{397}{nm} is applied in all schemes. Scheme I: \SI{866}{nm} laser applied to repump, \SI{397}{nm} fluorescence detected (no \SI{393}{nm} light emitted). Scheme II: \SI{850}{nm} and \SI{854}{nm} lasers applied, both violet wavelengths detected. Scheme FII: as scheme II but only \SI{393}{nm} light detected. Scheme FIII: all three infra-red lasers applied, only \SI{393}{nm} light detected.}
\label{fig:levelscheme}       
\end{figure}

Our Doppler cooling schemes are based on the standard $\lev{}{S}{1/2}\leftrightarrow \lev{}{P}{1/2}$ cooling transition which in \Ca{40} is at \SI{397}{nm} (see figure~\ref{fig:levelscheme}). The ion has a \SI{5}{\%} chance of 
decaying from the \lev{}{P}{1/2} level into the low-lying \lev{}{D}{3/2} level which is long-lived because the direct transition to the ground state is dipole-forbidden. To repump the ion, 
an infra-red laser is usually applied to the $\lev{}{D}{3/2}\leftrightarrow\lev{}{P}{1/2}$ transition at \SI{866}{nm}. This will be referred to as \textquotedblleft scheme I" here. 

We can instead repump via 
the \lev{}{P}{3/2} level using a laser at \SI{850}{nm}. This state decays quickly to the ground state by emitting a photon at \SI{393}{nm}. There is also a \SI{5}{\%} chance of decay into 
the \lev{}{D}{5/2} level from which we repump using a laser at \SI{854}{nm}. This repumping technique which we refer to as \textquotedblleft scheme II" makes the system into a quasi-two-level 
atom. It can add about 10\% of \SI{393}{nm} fluorescence to the signal and can increase the emitted \SI{397}{nm} light by about 70\% raising the overall signal by a substantial factor (see \cite{Allcock10} and table~\ref{T:fluo}), without increase in background. 

We now consider suppression of scattered laser light background (scheme FII). Detected light at \SI{393}{nm} originates exclusively from the ion which can be discriminated from the 
cooling light at \SI{397}{nm} and all the infra-red repumping lasers. The pair of interference filters is placed in front of the PMT to block selectively the cooling light while transmitting 
the \SI{393}{nm} light emitted by the ion.  This reduces background scatter from the cooling light by more than two orders of magnitude. We then only detect the ion's fluorescence, on a 
background level which is dominated by the dark counts of the PMT. All the advantages of standard Doppler cooling such as the cooling rate and the velocity capture range are maintained. The 
overall count rate is reduced however since \SI{397}{nm} light scattered by the ion is also discarded.

The detected fluorescence at 393\nm\ is limited by the $\lev{}{P}{1/2}\rightarrow\lev{}{D}{3/2}$ branching ratio (6.0\%) since only the population decaying to \lev{}{D}{3/2} can contribute. Scheme FIII addresses this disadvantage and achieves a higher fluorescence rate by actively pumping down more population from the \lev{}{P}{1/2} state with the usual repumping 
laser at \SI{866}{nm}. This increases the fluorescence rate at the cost of interconnecting several states which creates more possibilities 
for two-photon dark resonances \cite{Siemers92}. In the experimental implementation we varied the intensities and detunings of the infra-red lasers to maximize empirically the fluorescence. The detunings are not critical because at the optimum intensities the fluorescence peak is relatively broad. We achieved a 
count rate within a factor of two of the standard 397\nm+866\nm\ cooling scheme I. 

In order to account for these experimental values, we have implemented a full optical Bloch equations model. It takes into account all Zeeman sublevels and coherence effects. The \lev{}{P}{1/2} and \lev{}{P}{3/2} populations given by this model for our experimental parameters, and the corresponding fluorescence count rates, are compared in Table \ref{T:fluo} for the four repumping 
schemes described above. When calculating the count rates from the simulated populations, the angular distribution of the fluorescence from decay of the $\lev{}{P}{3/2}$ level was taken into account. The agreement between the predicted and measured rates is at the 5\% level. This is very satisfactory given that the empirical optimization procedure for the measured rates was only approximate. The theoretical results also depend on the direction and magnitude of the magnetic field which are uncertain because in addition to the applied field there is an unknown ambient component. A value of {\textbar{\bf{B}}\textbar}$\:=\:$\SI{3.8}{G} was chosen to give the highest level of consistency. This value is plausible given the presence of magnetic materials in the trap chip carrier, and our experience in other ion traps. For simplicity the direction was taken to be along that of the applied field. A net detection efficiency of 0.25\% at 397\nm\ was assumed; this is consistent both with the fluorescence spectrum in scheme I (fig. \ref{fig:scans_filterout}) and with an independent measurement using a pulsed method \cite{Shu10}. 

It would be possible to implement a similar set of schemes by cooling on the \SI{393}{nm} transition and repumping to the \lev{}{P}{1/2} level via the \SI{866}{nm} transition. Fluorescence at 
\SI{397}{nm} could then be collected and discriminated in a similar way. However, in the limit of full saturation of all transitions only the two states in \lev{}{P}{1/2} contribute to the 
fluorescence, compared with the four \lev{}{P}{3/2} states in our scheme, leading to a lower signal.

\begin{table*}[ht]
	\begin{center}
	\begin{tabular*}{7in}{|l|c|c|c|c|c||c|c|c|c|c|}
	\hhline{-----------}
	Scheme&\multicolumn{4}{c}{Laser powers $/ \mu$W}\vline&Filters&\multicolumn{2}{c}{Populations}\vline&\multicolumn{2}{c|}{Signal counts / s}&Background\\
	 &397&866&850&854& &P$_{1/2}$&P$_{3/2}$&predicted&measured&counts /s\\
	\hhline{-----------}
	I&22&9.9&--&--&out&0.231&0&77102&77800&990\\
	\hhline{-----------}
	II&22&--&37&6.6&out&0.357&0.022&125750&126000&990\\
	\hhline{-----------}
	FII&22&--&51&10&in&0.389&0.024&7651&7680&11\\
	\hhline{-----------}
	FIII&22&45&141&12&in&0.088&0.125&43643&41800&11\\
	\hhline{-----------}
	\end{tabular*}
	\end{center}
	\caption{Experimental parameters, measured fluorescence count rates and theoretical predictions based on the optical Bloch equations. For each repumping scheme the laser powers and detunings were empirically optimized to maximize the fluorescence signal. The columns give: the optimized laser powers; whether the filters were used or not; the expected total populations of the \lev{}{P}{1/2} and \lev{}{P}{3/2} levels for the laser intensities used in the experiments; the predicted and measured PMT count rates (excluding background); measured background count rates. Agreement between the predicted and measured rates is at the 5\% level (see text).}
	\label{T:fluo}
\end{table*}

\section{Experimental implementation}
\label{sec:exp}

\begin{figure}[ht]
\begin{center}
\resizebox{0.48\textwidth}{!}{\includegraphics{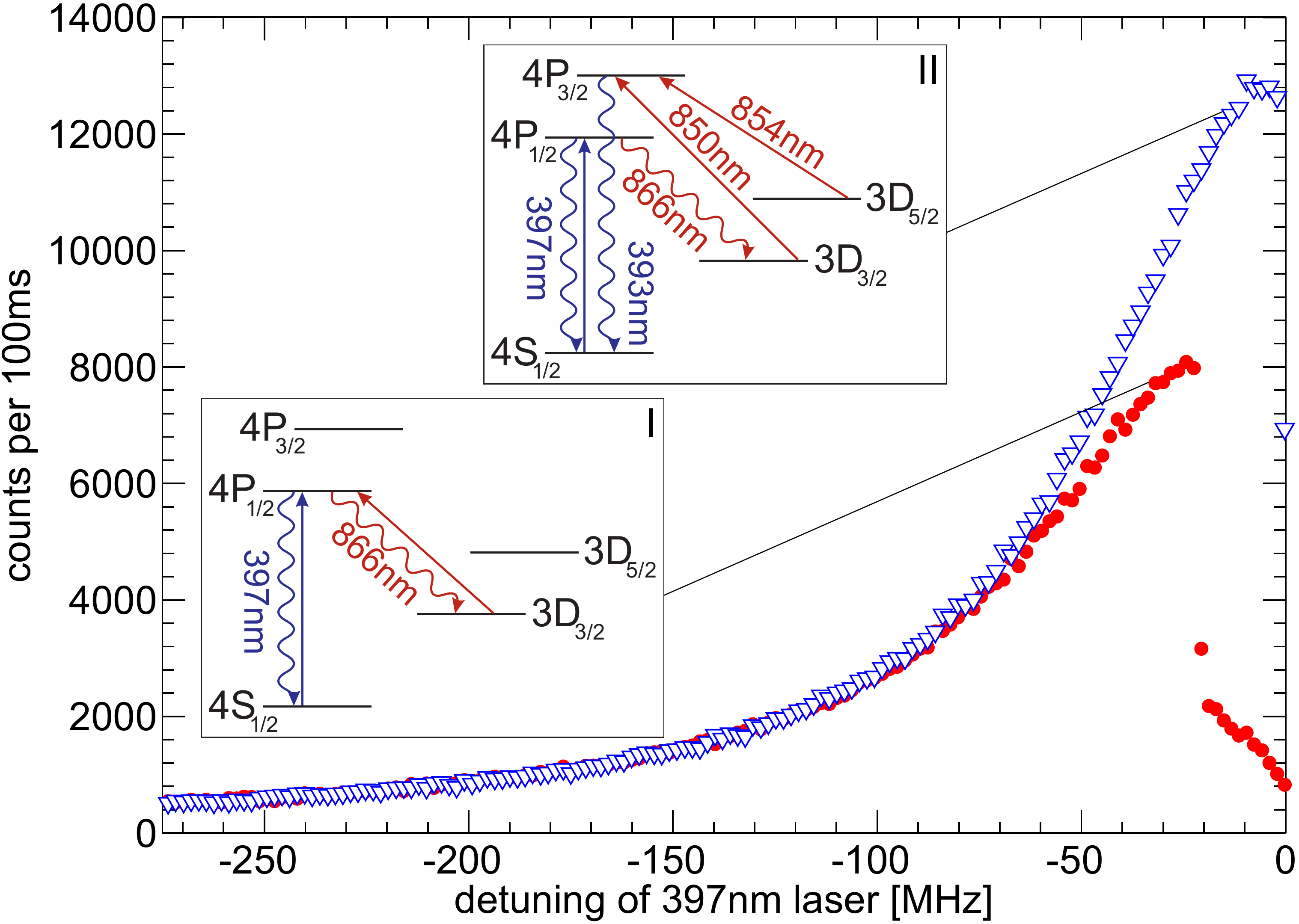}}
\end{center}
\caption{PMT count rate versus detuning of the laser at \SI{397}{nm}, for repumping schemes I and II. Both violet scattered wavelengths are detected. The 397\nm\ laser power was 22\uW. Zero 
detuning was inferred from the point at which the signal drops down sharply due to Doppler heating on the blue side of the resonance peak, in scheme II (in scheme I, this point occurs 
earlier because of a 397\nm/866\nm\ dark resonance). The insets show the two different repumping schemes. {\bf{I:}} repumping with a single infra-red laser at \SI{866}{nm}. {\bf{II:}} 
repumping with two lasers at \SI{850}{nm} and \SI{854}{nm}.}
\label{fig:scans_filterout}       
\end{figure}

In figure \ref{fig:scans_filterout} fluorescence spectra from a single ion are shown for schemes I and II. No filters were placed in front of the PMT, so both violet wavelengths can contribute to the signal. The maximum signals above background are 77800\persec and
126000\persec, respectively. 

\begin{figure}[t]
\begin{center}
\resizebox{0.48\textwidth}{!}{\includegraphics{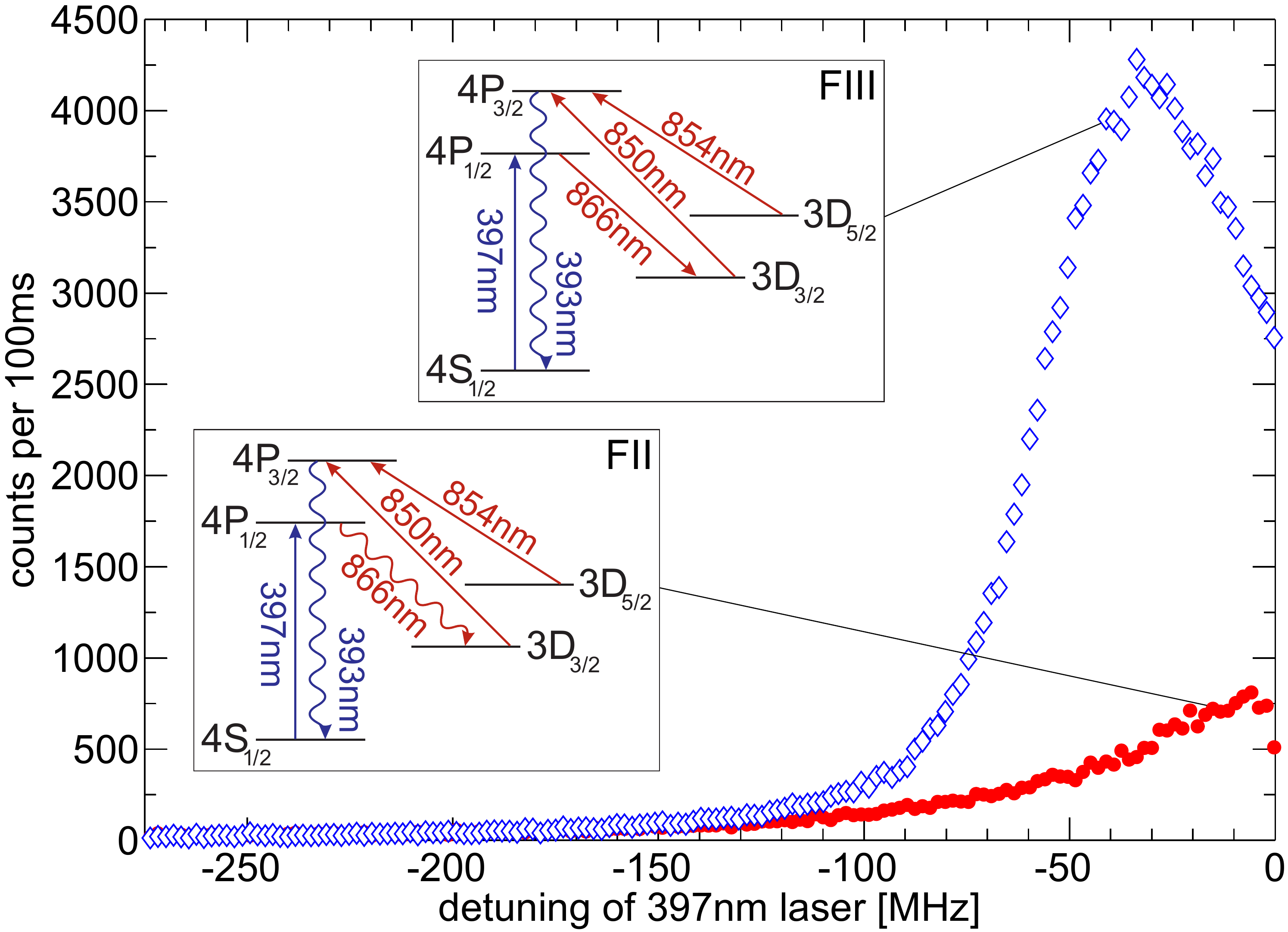}}
\end{center}
\caption{PMT count rate versus detuning of the laser at \SI{397}{nm}, for repumping schemes FII and FIII. The interference filters are placed in front of the PMT so that only scattered light 
at 393nm is detected. The 397\nm\ laser power was 22\uW. The insets show the two different repumping schemes. {\bf{FII:}} repumping with two lasers at \SI{850}{nm} and \SI{854}{nm}. {\bf{FIII:}} repumping with all three infra-red lasers to increase the fluorescing population in \lev{}{P}{3/2}. Again, the fluorescence peaks earlier in FIII because of a 397\nm/866\nm\ dark resonance.}
\label{fig:scans_filterin}       
\end{figure}

Placing the bandpass filters in front of the detector cuts out almost all light at \SI{397}{nm} and we only detect \SI{393}{nm} fluorescence. The results of implementing schemes FII and FIII are presented in figure \ref{fig:scans_filterin}. The ma\-xi\-mum signals are 7680\persec and 41800\persec\, respectively. Scheme FIII thus achieves a signal which is only a factor of two smaller than scheme I, but with the background reduced by two orders of magnitude.

\begin{figure}[t]
\begin{center}
\resizebox{0.48\textwidth}{!}{\includegraphics{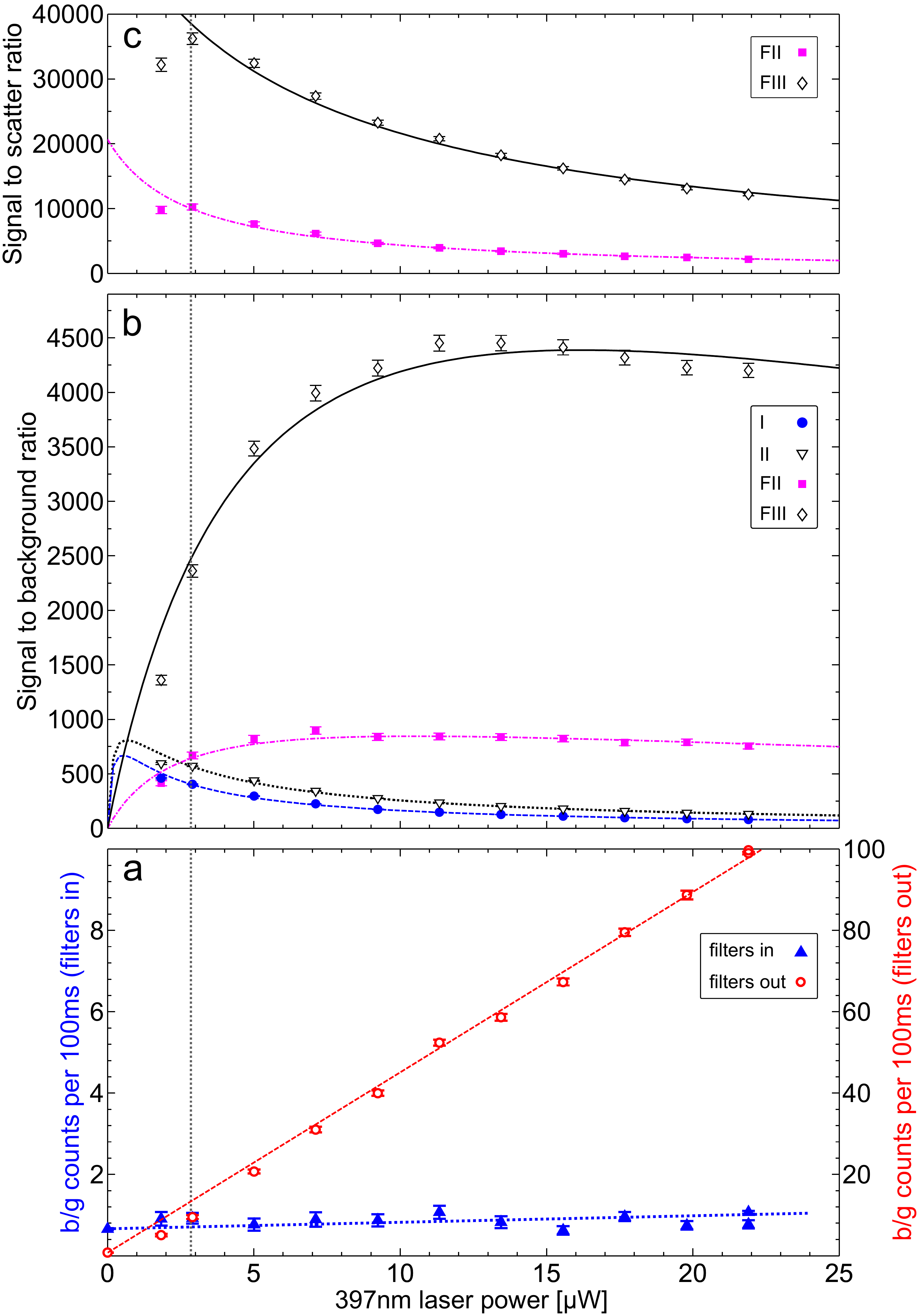}}
\end{center}
\caption{a: Background counts versus 397\nm\ laser power without the filters (open circles, right ordinate) 
and with the filters (filled triangles, left ordinate) including straight line fits. b: signal-to-background ratio versus 397\nm\ laser power for the four different repumping schemes, with fitted curves. Data points for the lowest 397 intensity are shifted down which may be due to a systematic reduction in the measured peak signal at very low intensities where the resonance becomes narrow and the peak fluorescence could be dropped, e.g. by finite laser linewidth. c: signal-to-scattered light ratio for the two schemes FII, FIII (for the schemes without the filters, this ratio is very similar to the signal-to-background ratio). The vertical dotted line shows the 397\nm\ laser power corresponding to the saturation intensity \Isat=2.8\,mW/mm$^2$.}
\label{fig:sigstobacks} 
\end{figure}

We repeated the four experiments presented in figures \ref{fig:scans_filterout} and \ref{fig:scans_filterin} for a range of different \SI{397}{nm} laser powers. At each power we took a data 
set with all infra-red beams turned off to determine the background which is shown in figure \ref{fig:sigstobacks}(a). The background is made up of the sum of the constant detector dark count rate and the rate of scatter which is proportional to the laser power. Without the filters the background rises steeply with 
increasing 397\nm\ power (slope 44.4(6)\persec$/\mu$W) while with the filters it is almost constant with a slope of only 0.16(2)\persec$/\mu$W). Straight lines are fitted to the two data 
sets to use as smooth background data for the signal-to-background comparison. The intercepts are 7(1)\persec\ and 6.6(4)\persec\ which corresponds to the PMT dark counts. 

We determined the peak fluorescence count for each $397\nm$ laser power $P_{397}$ and fitted a curve for the signal $S$ as a function of laser power $S=c/(1+s/P_{397})$, where $s$ and 
$c$ are constants. Although strictly appropriate only to a two-level system, we find that this formula is also in good agreement with calculations from our optical Bloch equations model. The four datasets are shown in figure \ref{fig:sigstobacks}(b) as signal-to-background points. The fitted curves, divided by the respective background slopes, 
are also shown. The standard scheme I has a peak signal-to-background ratio of 670. This is slightly improved by scheme II to 810. Both maxima are narrow peaks and occur at low laser 
intensity ($\ish 0.2\Isat$), where the reduced fluorescence leads to a smaller signal-to-noise ratio and lower cooling efficiency. With the filters in, scheme FII achieves a 
signal-to-background ratio of 850 which, while only slightly better than the previous results, is maintained over a wide range of 397\nm\ laser powers and is only gradually reduced in the 
limit of high laser power. The true merit of our technique can be seen in the results for scheme FIII: the signal-to-background ratio achieved is similar to the standard scheme at low intensities but then rises steeply and peaks at 4400. The peak is again very broad which makes the method insensitive to the laser intensity. At typical experimental parameters 
($P_{397}\approx 10\uW$) the method surpasses the standard 397\nm+866\nm\ scheme I by more than an order of magnitude. 

Since with the filters, the background is dominated by the dark counts, which are highly detector-dependent, it is important to consider also the ratio of signal to scattered laser light. 
This is shown for the two schemes with the filters in figure~\ref{fig:sigstobacks}(c), and has a maximum of 36000 for the scheme FIII. This occurs at a 397\nm\ laser power 
corresponding to approximately one saturation intensity (2.8\,mW/mm$^2$), which offers a good compromise between high scattering rate and low Doppler cooling temperature. For a different 
detector, such as a cooled EMCCD at the same 100\ms\ integration time, the dark counts can be negligible and the signal-to-background achievable should be close to this signal-to-scatter 
ratio.

\section{Applications and outlook}
\label{sec:advantages}

Wherever ions are trapped close to surfaces and scattered light from the laser cooling beam is an issue, this technique can be implemented to increase significantly the signal-to-background 
ratio. This is the case in miniaturized scalable trap geometries for QIP as well as other recent experiments in ion traps. Apart from the cavity and fibre applications mentioned in the 
introduction, trapped ions are used in sensor applications \cite{Guthorlein01} employing them as field probes in specifically designed traps \cite{Maiwald09}. Other areas of study 
are ion-surface interactions and surface plasmon experiments in nano-wires \cite{BrownnuttSelber}. In all these cases, the ease of implementation with only broadly stabilized diode lasers as 
well as its ready applicability to other ion species with similar level structures make this scheme attractive. 

A low background level is important not only for ion detection, but also for
accurate compensation of ion micromotion (for example, by the standard
technique of rf/photon arrival time correlation \cite{Berkeland98}), or for thermometry as recently described in \cite{Norton11}. Our
scheme can also be used in conjunction with heating rate measurements
by Doppler recooling as described in \cite{Wesenberg07}. The theory given in
\cite{Wesenberg07} is based on a two-level system which is not realized
exactly in the standard repumping scheme I involving 3 levels  connected in
a $\mathrm{\Lambda}$ arrangement. Repumping with the 850\nm\ and 854\nm\
lasers (scheme II) gives a close approximation to a two-level system, as
discussed in \cite{Allcock10}, which allows simple interpretation of the
Doppler recooling data to extract the heating rate. In cases where there is
high background scatter, scheme FII can be employed. \newline

\noindent {\bf Acknowledgements} We are extremely grateful to Sandia National Laboratories for their fabrication of the trap. The optical Bloch equation simulations were implemented using NAG software. We would also like to acknowledge Alice Burrell for the construction 
of the imaging system, Graham Quelch for laboratory support and Luca Guidoni for comments on the manuscript. This work was supported by the EPSRC Science and Innovation programme.

%
%
%
%
 \bibliographystyle{unsrt}
 \bibliography{LitJo}
%
%
%

\end{document}